\documentclass[preprint,showpacs,superscriptaddress,nofootinbib]{revtex4}
  \usepackage{graphicx}
  \usepackage{rotating}
  \usepackage{multirow}
  \begin{document}
  \title{Study on ${\Upsilon}(nS)$ ${\to}$ $B_{c}M$ decays}
  \author{Junfeng Sun}
  \affiliation{Institute of Particle and Nuclear Physics,
              Henan Normal University, Xinxiang 453007, China}
  \author{Lili Chen}
  \affiliation{Institute of Particle and Nuclear Physics,
              Henan Normal University, Xinxiang 453007, China}
  \author{Na Wang}
  \affiliation{Institute of Particle and Key Laboratory of Quark and Lepton Physics,
              Central China Normal University, Wuhan 430079, China}
  \author{Jinshu Huang}
  \affiliation{College of Physics and Electronic Engineering,
              Nanyang Normal University, Nanyang 473061, China}
  \author{Yueling Yang}
  \affiliation{Institute of Particle and Nuclear Physics,
              Henan Normal University, Xinxiang 453007, China}
  \author{Qin Chang}
  \affiliation{Institute of Particle and Nuclear Physics,
              Henan Normal University, Xinxiang 453007, China}

  \begin{abstract}
  With anticipation of abundant Upsilons data sample at
  high-luminosity heavy-flavor experiments in the future,
  we studied nonleptonic two-body weak decays of
  ${\Upsilon}(nS)$ below the open-bottom threshold with
  $n$ $=$ $1$, $2$ and $3$.
  It is found that branching ratios for ${\Upsilon}(1S,2S,3S)$
  ${\to}$ $B_{c}{\rho}$ decays are relatively large among
  Upsilons decay into $B_{c}M$ final states ($M$ $=$ ${\pi}$
  ${\rho}$, $K$ and $K^{\ast}$) and can reach up to
  $10^{-10}$,
  which is promisingly detected by experiments at the
  running LHC and forthcoming SuperKEKB.
  \end{abstract}
  \maketitle

  \section{Introduction}
  \label{sec01}
  About forty years after the discovery of Upsilons (the bound
  states of $b\bar{b}$ with quantum number of $I^{G}J^{PC}$ $=$
  $0^{-}1^{--}$ \cite{pdg}) at Fermilab in 1977 \cite{bb1977},
  the properties of bottomonium system continue to be the subject
  of intensive theoretical and experimental study.
  Major contributions were made recently by experiments at the
  asymmetric electron-positron colliders KEK-B with Belle detector
  and PEP-II with BaBar detector, and the hadron colliders
  Tevatron and LHC \cite{1212.6552}.

  Some of the salient features of Upsilons are as
  follows \cite{ann1983}:
  (1)
  In the center-of-mass frame of Upsilons, the relative
  motion of the bottom quark is sufficiently slow.
  Nonrelativistic Schr\"{o}dinger equation can be used
  to describe well the spectrum of bottomonium system
  and thus one can learn about the interquark binding
  forces.
  (2)
  The ${\Upsilon}(nS)$ particles below the open-bottom
  threshold, with the radial quantum number $n$ $=$ $1$,
  $2$ and $3$, decay primarily via the annihilation of
  the $b\bar{b}$ quark pairs into three gluons, which
  also provide an entry to many potential final states
  including glueballs, hybrid and multiquark states.
  Thus the properties of the invisible gluons and of the
  gluon-quark coupling can be gleaned through the study
  of hadronic Upsilons decay.
  (3)
  Compared with the light $u$, $d$, $s$ quarks,
  the relatively large mass of the $b$ quark
  implies a nonnegligible coupling to the Higgs bosons,
  making Upsilons to be one of the best hunting
  grounds for light Higgs particles.
  By now, our knowledge of the properties of Upsilons
  comes mostly from $e^{+}e^{-}$ collision.

  As is well known, Upsilons decay mainly through the
  strong and electromagnetic interactions.
  The coupling constant ${\alpha}_{s}$ for hadronic Upsilons
  decay is smaller than that for charmonium decay due to the
  Quantum Chromodynamics (QCD) asymptotic freedom.
  In addition, the coupling between Upsilons and photon is
  proportional to the electric charges of the bottom quark.
  So, one of the outstanding properties of Upsilons below
  $B\bar{B}$ threshold is their narrow decay width of tens
  of keV \cite{pdg} (see Table \ref{tab:bb}).
  Besides, as an essential complement to Upsilons decay modes,
  the Upsilons weak decay is allowable within the standard
  model and might be accessible at experiments,
  although the branching ratio is tiny, about
  $2/{\tau}_{B}{\Gamma}_{\Upsilon}$
  ${\sim}$ $10^{-8}$ \cite{pdg}.
  In this paper, we will estimate the branching ratios
  for nonleptonic two-body ${\Upsilon}(nS)$ ${\to}$
  $B_{c}M$ weak decays, where $M$ $=$ ${\pi}$, ${\rho}$,
  $K$ and $K^{\ast}$.
  The motivation is listed as follows.

   \begin{table}[h]
   \caption{Summary of the mass, decay width and data samples
   of Upsilons below $B\bar{B}$ threshold collected by Belle,
   BaBar and CLEO Collaborations.}
   \label{tab:bb}
   \begin{ruledtabular}
   \begin{tabular}{c|c|c|c|c|c}
    & \multicolumn{2}{c|}{properties \cite{pdg}}
    & \multicolumn{3}{c}{data samples ($10^{6}$)} \\ \cline{2-6}
      meson & mass (MeV) & width (keV)
    & Belle \cite{1212.5342}
    & BaBar \cite{1108.5874}
    & CLEO  \cite{0704.2766} \\ \hline
   ${\Upsilon}(1S)$ & $ 9460.30{\pm}0.26$ & $54.02{\pm}1.25$ & 102 & ...   & 22.78  \\
   ${\Upsilon}(2S)$ & $10023.26{\pm}0.31$ & $31.98{\pm}2.63$ & 158 & 121.8 &  9.45 \\
   ${\Upsilon}(3S)$ & $10355.2 {\pm}0.5 $ & $20.32{\pm}1.85$ &  11 &  98.6 &  8.89
   \end{tabular}
   \end{ruledtabular}
   \end{table}

  From the experimental point of view,
  (1)
  there is plenty of Upsilons at the high-luminosity
  dedicated bottomonium factories.
  Over $10^{8}$ Upsilons data samples have been collected
  at Belle and BaBar experiments (see Table \ref{tab:bb}).
  Upsilons are also observed by the on-duty
  ALICE \cite{1403.3648}, ATLAS \cite{1212.7255},
  CMS \cite{1501.07750}, LHCb \cite{1304.6977}
  experiments at LHC.
  It is hopefully expected that more than $10^{11}$ $b\bar{b}$
  quark pairs would be available per $fb^{-1}$ data
  at LHCb \cite{1408.0403} and huge Upsilons data samples
  could be accumulated with great precision at the
  forthcoming SuperKEKB \cite{superb}.
  A large amount of data samples will provide opportunities
  to search for Upsilons weak decays which in some cases
  might be detectable.
  Hence, theoretical studies on Upsilons weak decays are
  very necessary to offer a ready reference.
  (2)
  For nonleptonic two-body ${\Upsilon}(nS)$ ${\to}$
  $B_{c}M$ weak decay,
  the final states with opposite charges have definite
  energy and momentum in the ${\Upsilon}(nS)$ rest frame.
  In addition, identification of a single charged
  $B_{c}$ meson would provide an unambiguous signature
  of Upsilons weak decay, which is free from double
  tagging of the $b$-flavored hadron pairs.
  The small branching ratios make the
  observation of Upsilons weak decays very difficult,
  and evidences of an abnormally production rate of a
  single $B_{c}$ meson in Upsilons decay might be
  a hint of new physics beyond the standard model.

  From the theoretical point of view, nonleptonic
  Upsilon weak decay could allow one to overconstrain
  parameters obtained from $B$ meson decay,
  test various models and improve our understanding
  on the strong interactions and the mechanism responsible
  for heavy meson weak decay.
  Phenomenologically, the ${\Upsilon}(nS)$ ${\to}$ $B_{c}M$
  weak decays are monopolized by tree contribution and favored
  by the Cabibbo-Kobayashi-Maskawa (CKM) quark mixing matrix
  element $V_{cb}$, so they should have relatively large
  branching ratios.
  The amplitudes for the ${\Upsilon}(nS)$
  ${\to}$ $B_{c}M$ decay are commonly written as factorizable
  product of two factors: one describing the
  transition between Upsilon and $B_{c}$ meson, and the
  other depicting the production of the $M$ state
  from the vacuum. The earlier works,
  including Refs. \cite{zpc62,ijma14} based on a heavy
  quark effective theory and
  Ref. \cite{adv2013} based on the Bauer-Stech-Wirbel
  (BSW) model \cite{bsw},
  riveted mainly upon the ${\Upsilon}(1S)$ ${\to}$ $B_{c}$
  transition form factors.
  No research works devoted to nonleptonic ${\Upsilon}(2S)$,
  ${\Upsilon}(3S)$ weak decays.
  In recent years, several attractive QCD-inspired
  methods have been developed to treat with the hadronic
  matrix elements of heavy flavor weak decay.
  In this paper, we will estimate branching ratios for
  nonleptonic two-body ${\Upsilon}(nS)$ ${\to}$ $B_{c}M$
  weak decay,
  by considering nonfactorizable contributions
  to hadronic matrix elements
  with the QCD factorization (QCDF) approach \cite{qcdf},
  and calculating the transition form factor between
  Upsilon and $B_{c}$ meson with nonrelativistic
  wave functions.

  This paper is organized as follows.
  In section \ref{sec02}, we will present the theoretical framework
  and the amplitudes for ${\Upsilon}(nS)$ ${\to}$ $B_{c}M$ decays.
  Section \ref{sec03} is devoted to numerical results and discussion.
  The last section is our summary.

  \section{theoretical framework}
  \label{sec02}
  \subsection{The effective Hamiltonian}
  \label{sec0201}
  Using the operator product expansion technique, the
  effective Hamiltonian responsible for ${\Upsilon}(nS)$
  ${\to}$ $B_{c}M$ decays is \cite{9512380}
   \begin{equation}
  {\cal H}_{\rm eff}\ =\ \frac{G_{F}}{\sqrt{2}}\,
   \sum\limits_{q=d,s}\, V_{cb} V_{uq}^{\ast}\,
   \Big\{ C_{1}({\mu})\,Q_{1}({\mu})
         +C_{2}({\mu})\,Q_{2}({\mu}) \Big\}
   + {\rm h.c.}
   \label{hamilton},
   \end{equation}
  where the Fermi coupling constant $G_{F}$ $=$
  $1.166{\times}10^{-5}\,{\rm GeV}^{-2}$ \cite{pdg};
  The CKM factors can be expanded as a power series in
  the Wolfenstein parameter ${\lambda}$ $=$ $0.22537(61)$
  \cite{pdg},
  \begin{eqnarray}
  V_{cb}V_{ud}^{\ast} &=&
               A{\lambda}^{2}
  - \frac{1}{2}A{\lambda}^{4}
  - \frac{1}{8}A{\lambda}^{6}
  +{\cal O}({\lambda}^{8})
  \label{eq:ckm01}, \\
  V_{cb}V_{us}^{\ast} &=& A{\lambda}^{3}
  +{\cal O}({\lambda}^{8})
  \label{eq:ckm02}.
  \end{eqnarray}

  The Wilson coefficients $C_{1,2}(\mu)$ summarize the
  physical contributions above scales of ${\mu}$,
  which are calculable with the perturbation theory
  and have properly been evaluated to the
  next-to-leading order (NLO).
  Their values at scale of ${\mu}$ ${\sim}$
  ${\cal O}(m_{b})$ can be evaluated with the
  renormalization group (RG) equation \cite{9512380}
  \begin{equation}
  C_{1,2}({\mu}) = U({\mu},m_{W})C_{1,2}(m_{W})
  \label{ci},
  \end{equation}
  where the RG evolution matrix $U({\mu},m_{W})$
  transforms the Wilson coefficients from scale of $m_{W}$
  to ${\mu}$.
  The expression of $U({\mu},m_{W})$ can be found
  in Ref. \cite{9512380}.
  With the naive dimensional regularization (NDR) scheme,
  the numerical values of Wilson coefficients $C_{1,2}$
  are listed in Table \ref{tab:ci}.

  The local tree four-quark operators are defined as follows.
    \begin{eqnarray}
    Q_{1} &=&
  [ \bar{c}_{\alpha}{\gamma}_{\mu}(1-{\gamma}_{5})b_{\alpha} ]
  [ \bar{q}_{\beta} {\gamma}^{\mu}(1-{\gamma}_{5})u_{\beta} ]
    \label{q1}, \\
    Q_{2} &=&
  [ \bar{c}_{\alpha}{\gamma}_{\mu}(1-{\gamma}_{5})b_{\beta} ]
  [ \bar{q}_{\beta}{\gamma}^{\mu}(1-{\gamma}_{5})u_{\alpha} ]
    \label{q2},
    \end{eqnarray}
  where ${\alpha}$ and ${\beta}$ are color indices and the
  sum over repeated indices is understood.

  To obtain the decay amplitudes, the remaining and also
  the most intricate part is how to calculate accurately
  hadronic matrix elements squeezing the local operators
  between initial Upsilons and final $B_{c}M$ states.

  \subsection{Hadronic matrix elements}
  \label{sec0202}
  Analogous to the usual applications of hard exclusive
  processes in perturbative QCD proposed by Lepage and
  Brodsky \cite{prd22},
  the QCDF approach is based on the collinear factorization
  approximation and power countering rules in the heavy quark
  limit, where hadronic matrix elements are written
  as the convolution integrals of hard scattering
  subamplitudes and universal wave functions \cite{qcdf}.
  The QCDF approach has been widely applied to $B$
  meson weak decays.
  As for the ${\Upsilon}(nS)$ ${\to}$ $B_{c}M$ decay,
  using the QCDF master formula, hadronic matrix elements
  can be written as :
   \begin{equation}
  {\langle}B_{c}M{\vert}Q_{i}{\vert}{\Upsilon}{\rangle} =
   \sum\limits_{j} F_{j}^{ {\Upsilon}{\to}B_{c} }
  {\int}\,dx\, H_{j}(x)\,{\Phi}_{M}(x)
   \label{hadronic},
   \end{equation}
  where both transition form factor $F_{i}^{ {\Upsilon}{\to}B_{c} }$
  and wave function ${\Phi}_{M}(x)$ are universal and
  nonperturbative input parameters.
  For a light pseudoscalar and vector meson,
  the leading twist distribution amplitude can be expressed
  in terms of Gegenbauer polynomials \cite{ball}:
   \begin{equation}
  {\phi}_{M}(x)=6\,x\bar{x}
   \sum\limits_{n=0}^{\infty}
   a_{n}^{M}\, C_{n}^{3/2}(x-\bar{x})
   \label{twist},
   \end{equation}
  where $\bar{x}$ $=$ $1$ $-$ $x$;
  $a_{n}^{M}$ is the Gegenbauer moment
  and $a_{0}^{M}$ ${\equiv}$ $1$.

  Hard scattering function, $H_{j}(x)$, is assumed to be
  calculable order by order from the first principle
  of perturbative QCD theory.
  At order of ${\alpha}_{s}^{0}$, $H_{j}(x)$ $=$ $1$
  and the integral for wave function in Eq.(\ref{hadronic})
  results in decay constant $f_{M}$, which is the simplest
  scenario.
  At order of ${\alpha}_{s}$ and higher orders, expression
  of $H_{j}(x)$ is no longer trivial,
  part of strong phases and renormalization scale dependence
  of amplitude can be recuperated from hadronic matrix
  elements.
  The decay amplitudes could be written as
   \begin{equation}
  {\cal A}({\Upsilon}{\to}B_{c}M) =
  {\langle}B_{c}M{\vert}{\cal H}_{\rm eff}{\vert}{\Upsilon}{\rangle} =
   \frac{G_{F}}{\sqrt{2}}\, V_{cb} V_{uq}^{\ast}\, a_{1}\,
  {\langle}M{\vert}J^{\mu}{\vert}0{\rangle}
  {\langle}B_{c}{\vert}J_{\mu}{\vert}{\Upsilon}{\rangle}
   \label{amp}.
   \end{equation}

  The coefficient $a_{1}$ in Eq.(\ref{amp}),
  containing nonfactorizable contributions to hadronic
  matrix elements, is written as:
  \begin{equation}
   a_{1}
    = C_{1}^{\rm NLO}+\frac{1}{N_{c}}\,C_{2}^{\rm NLO}
    + \frac{{\alpha}_{s}}{4{\pi}}\, \frac{C_{F}}{N_{c}}\,
      C_{2}^{\rm LO}\, V
   \label{a1}.
  \end{equation}
  The explicit expression of parameter $V$ is the same
  as that in Ref. \cite{prd77}.
  It has been shown that coefficient $a_{1}$ is infrared-safe
  and renormalization scale independent at order of
  ${\alpha}_{s}$ \cite{prd77}.
  The numerical values of coefficient $a_{1}$
  at scales of ${\mu}$ ${\sim}$
  ${\cal O}(m_{b})$ are listed in Table \ref{tab:ci}.
  From the numbers in Table \ref{tab:ci}, it is seen
  that one could get part information of strong phase
  by taking nonfactorizable corrections into account,
  though the strong phase is small and
  suppressed by factor ${\alpha}_{s}/N_{c}$.

   \begin{table}[h]
   \caption{Numerical values of the Wilson coefficients $C_{1,2}$
    with NDR scheme and parameter $a_{1}$ for the ${\Upsilon}(nS)$
    ${\to}$
    $B_{c}{\pi}$ decay, where $m_{b}$ $=$ 4.78 GeV \cite{pdg}.}
   \label{tab:ci}
   \begin{ruledtabular}
   \begin{tabular}{cccccc}
 & \multicolumn{2}{c}{LO} & \multicolumn{2}{c}{NLO} & \\ \cline{2-3} \cline{4-5}
 ${\mu}$ & $C_{1}$ & $C_{2}$ & $C_{1}$ & $C_{2}$ & $a_{1}$ \\ \hline
 $0.5\,m_{b}$ &
 $1.173$ & $-0.346$ &
 $1.132$ & $-0.277$ &
 $1.081e^{-i2^{\circ}}$ \\
 $m_{b}$ &
 $1.112$ & $-0.240$ &
 $1.078$ & $-0.176$ &
 $1.057e^{-i1^{\circ}}$ \\
 $1.5\,m_{b}$ &
 $1.086$ & $-0.192$ &
 $1.055$ & $-0.130$ &
 $1.045e^{-i1^{\circ}}$ \\
 $2.0\,m_{b}$ &
 $1.071$ & $-0.161$ &
 $1.042$ & $-0.101$ &
 $1.038e^{-i1^{\circ}}$
   \end{tabular}
   \end{ruledtabular}
   \end{table}

  \subsection{Decay constants and form factors}
  \label{sec0203}
  The matrix elements of current operators are defined as follows:
   \begin{eqnarray}
  {\langle}P(p){\vert}A_{\mu}{\vert}0{\rangle}
  &=&
   -if_{P}\,p_{\mu}
   \label{cme01}, \\
  {\langle}V(p,{\epsilon}){\vert}V_{\mu}{\vert}0{\rangle}
  &=&
   f_{V}\,m_{V}\,{\epsilon}_{V,{\mu}}^{\ast}
   \label{cme02},
   \end{eqnarray}
 where $f_{P}$ and $f_{V}$ are the decay constants
 of pseudoscalar and vector mesons, respectively;
 $m_{V}$ and ${\epsilon}_{V}$ denote the mass and
 polarization of vector meson, respectively.

  The transition form factors are defined as follows
  \cite{ijma14,adv2013,bsw}:
    \begin{eqnarray}
   & &
   {\langle}B_{c}(p_{2}){\vert}V_{\mu}-A_{\mu}
   {\vert}{\Upsilon}(p_{1},{\epsilon}){\rangle}
    \nonumber \\ &=&
  -{\epsilon}_{{\mu}{\nu}{\alpha}{\beta}}\,
   {\epsilon}_{{\Upsilon}}^{{\nu}}\,
    q^{\alpha}\, (p_{1}+p_{2})^{\beta}\,
     \frac{V^{{\Upsilon}{\to}B_{c}}(q^{2})}{m_{{\Upsilon}}+m_{B_{c}}}
   -i\,\frac{2\,m_{{\Upsilon}}\,{\epsilon}_{{\Upsilon}}{\cdot}q}{q^{2}}\,
    q_{\mu}\, A_{0}^{{\Upsilon}{\to}B_{c}}(q^{2})
    \nonumber \\ & &
    -i\,{\epsilon}_{{\Upsilon},{\mu}}\,
    ( m_{{\Upsilon}}+m_{B_{c}} )\, A_{1}^{{\Upsilon}{\to}B_{c}}(q^{2})
   -i\,\frac{{\epsilon}_{{\Upsilon}}{\cdot}q}{m_{{\Upsilon}}+m_{B_{c}}}\,
   ( p_{1} + p_{2} )_{\mu}\, A_{2}^{{\Upsilon}{\to}B_{c}}(q^{2})
    \nonumber \\ & &
   +i\,\frac{2\,m_{{\Upsilon}}\,{\epsilon}_{{\Upsilon}}{\cdot}q}{q^{2}}\,
   q_{\mu}\, A_{3}^{{\Upsilon}{\to}B_{c}}(q^{2})
    \label{cme03},
    \end{eqnarray}
  where $q$ $=$ $p_{1}$ $-$ $p_{2}$;
  and $A_{0}(0)$ $=$ $A_{3}(0)$
  is required compulsorily to cancel singularities at the
  pole $q^{2}$ $=$ $0$.
  There is a relation among these form factors
  \begin{equation}
   2m_{{\Upsilon}}A_{3}(q^{2})=
   (m_{{\Upsilon}}+m_{B_{c}})A_{1}(q^{2})
  +(m_{{\Upsilon}}-m_{B_{c}})A_{2}(q^{2})
  \label{form01}.
  \end{equation}

  The form factors, $A_{0,1}(0)$
  and $V(0)$ at the pole $q^{2}$ $=$ $0$ are defined
  as \cite{bsw},
  \begin{equation}
  A_{0}^{{\Upsilon}{\to}B_{c}}(0) =
  {\int}d\vec{k}_{\perp} {\int}_{0}^{1}dx\,
   \Big\{ {\Phi}_{\Upsilon}(\vec{k}_{\perp},x,1,0)\,
  {\sigma}_{z}\, {\Phi}_{B_{c}}(\vec{k}_{\perp},x,0,0) \Big\}
  \label{form-a0},
  \end{equation}
  \begin{eqnarray}
  A_{1}^{ {\Upsilon}{\to}B_{c} }(0) &=&
  \frac{ m_{b}+m_{c} }{ m_{{\Upsilon}}+m_{B_{c}} }
  I^{{\Upsilon}{\to}B_{c}}
  \label{form-a1} \\
  V^{ {\Upsilon}{\to}B_{c} }(0)  &=&
  \frac{ m_{b}-m_{c} }{ m_{{\Upsilon}}-m_{B_{c}} }
  I^{{\Upsilon}{\to}B_{c}}
  \label{form-v},
  \end{eqnarray}
  \begin{equation}
  I^{{\Upsilon}{\to}B_{c}} = \sqrt{2}
  {\int}d\vec{k}_{\perp} {\int}_{0}^{1} \frac{dx}{x}\,
   \Big\{ {\Phi}_{\Upsilon}(\vec{k}_{\perp},x,1,-1)\,
  i{\sigma}_{y}\, {\Phi}_{B_{c}}(\vec{k}_{\perp},x,0,0) \Big\}
  \label{form-ii},
  \end{equation}
  where ${\sigma}_{y,z}$ is a Pauli matrix acting on
  the spin indices of the decaying bottom quark;
  $x$ and $\vec{k}_{\perp}$ denote the fraction of
  the longitudinal momentum and the transverse momentum
  carried by the nonspectator quark, respectively.

  For Upsilons, the bottom quark is nonrelativistic with
  an average velocity $v$ ${\ll}$ $1$ based on arguments
  of nonrelativistic quantum chromodynamics
  (NRQCD) \cite{qcd}.
  For the double-heavy $B_{c}$ meson, both bottom and charm
  quarks are nonrelativistic due to $m_{B_{c}}$ ${\approx}$
  $m_{b}$ $+$ $m_{c}$.
  Here, we will take the solution of the Sch\"{o}dinger equation
  with an isotropic harmonic oscillator potential as wave
  functions of Upsilons and $B_{c}$ states, i.e.,
   \begin{equation}
  {\phi}_{1S}(\vec{k})\ {\sim}\
   e^{-\vec{k}^{2}/2{\alpha}^{2}}
   \label{wave-r1s},
   \end{equation}
   \begin{equation}
  {\phi}_{2S}(\vec{k})\ {\sim}\
   e^{-\vec{k}^{2}/2{\alpha}^{2}}
   ( 2\vec{k}^{2}-3{\alpha}^{2} )
   \label{wave-r2s},
   \end{equation}
   \begin{equation}
  {\phi}_{3S}(\vec{k})\ {\sim}\
   e^{-\vec{k}^{2}/2{\alpha}^{2}}
   ( 4\vec{k}^{4}-20\vec{k}^{2}{\alpha}^{2}+15{\alpha}^{4} )
   \label{wave-r3s},
   \end{equation}
  where the parameter ${\alpha}$ determines the average
  transverse quark momentum,
  ${\langle}{\phi}_{1S}{\vert}\vec{k}^{2}_{\perp}
   {\vert}{\phi}_{1S}{\rangle}$ $=$ ${\alpha}^{2}$.
  With the NRQCD power counting rules \cite{qcd},
  ${\vert}\vec{k}_{\perp}{\vert}$ ${\sim}$ $mv$ ${\sim}$
  $m{\alpha}_{s}$ for heavy quarkonium.
  Hence, parameter ${\alpha}$ is approximately taken as
  $m{\alpha}_{s}$ in our calculation.
  Using the substitution ansatz \cite{xiao},
   \begin{equation}
   \vec{k}^{2}\ {\to}\
   \frac{\vec{k}_{\perp}^{2}+\bar{x}\,m_{q}^{2}+x\,m_{b}^{2}}{4\,x\,\bar{x}}
   \label{wave04},
   \end{equation}
   one can obtain
   \begin{equation}
  {\phi}_{1S}(\vec{k}_{\perp},x) = A\,
  {\exp}\Big\{ \frac{\vec{k}_{\perp}^{2}+\bar{x}\,m_{q}^{2}+x\,m_{b}^{2}}
                    {-8\,{\alpha}^{2}\,x\,\bar{x}} \Big\}
   \label{wave-k1s},
   \end{equation}
   \begin{equation}
  {\phi}_{2S}(\vec{k}_{\perp},x) = B\, {\phi}_{1S}(\vec{k}_{\perp},x)
   \Big\{ \frac{\vec{k}_{\perp}^{2}+m_{b}^{2}}
                    {6\,{\alpha}^{2}\,x\,\bar{x}}-1 \Big\}
   \label{wave-k2s},
   \end{equation}
   \begin{equation}
  {\phi}_{3S}(\vec{k}_{\perp},x) = C\, {\phi}_{1S}(\vec{k}_{\perp},x)
   \Big\{ \frac{2}{5}\Big( \frac{\vec{k}_{\perp}^{2}+m_{b}^{2}}
                     {4\,{\alpha}^{2}\,x\,\bar{x}}-\frac{5}{2} \Big)^2 -1 \Big\}
   \label{wave-k3s},
   \end{equation}
   where parameters $A$, $B$ and $C$ are normalization factors.

   \begin{table}[h]
   \caption{Numerical values of transition form factors at
    $q^{2}$ $=$ $0$, where uncertainties of this work are
    from the masses of bottom and charm quarks, and numbers
    in Ref. \cite{adv2013} are computed with the flavor
    dependent parameter ${\omega}$ the BSW model.}
   \label{tab:ff}
   \begin{ruledtabular}
   \begin{tabular}{c|c|cccc}
   transition & reference
 & $A_{0}(0)$ & $A_{1}(0)$ & $A_{2}(0)$ & $V(0)$ \\ \hline
   ${\Upsilon}(1S)$ ${\to}$ $B_{c}$
 & \cite{adv2013}
 & $0.46$
 & $0.62$
 & $0.38$
 & $1.61$ \\
 & this work
 & $0.67{\pm}0.02$
 & $0.70{\pm}0.02$
 & $0.51{\pm}0.06$
 & $1.66{\pm}0.02$ \\ \hline
   ${\Upsilon}(2S)$ ${\to}$ $B_{c}$
 & this work
 & $0.65{\pm}0.02$
 & $0.69{\pm}0.02$
 & $0.48{\pm}0.04$
 & $1.44{\pm}0.03$ \\ \hline
   ${\Upsilon}(3S)$ ${\to}$ $B_{c}$
 & this work
 & $0.57{\pm}0.01$
 & $0.64{\pm}0.01$
 & $0.29{\pm}0.03$
 & $1.25{\pm}0.05$
   \end{tabular}
   \end{ruledtabular}
   \end{table}

   The numerical values of transition form factors at $q^{2}$ $=$ $0$
   are collected in Table \ref{tab:ff}.
   It is found that (1) form factors for the ${\Upsilon}(1S)$ ${\to}$
   $B_{c}$ transition are generally larger than
   those in Ref. \cite{adv2013}.
   (2) The value of form factor at $q^{2}$ $=$ $0$ decreases gradually
   with the increase of the radial quantum number of Upsilons.

  \subsection{Decay amplitudes}
  \label{sec0204}
  With the above definition of hadronic matrix elements,
  the decay amplitudes for ${\Upsilon}(nS)$ ${\to}$ $B_{c}M$
  decays can be written as
    \begin{eqnarray}
  {\cal A}({\Upsilon}{\to}B_{c}^{+}{\pi}^{-})
  &=&
   \sqrt{2}\, G_{F}\, V_{cb}\, V_{ud}^{\ast}\, a_{1}\,
   f_{\pi}\, m_{\Upsilon}\, ({\epsilon}_{\Upsilon}{\cdot}p_{\pi})\,
    A_{0}^{{\Upsilon}{\to}B_{c}}
   \label{amp-bc-pi}, \\
  {\cal A}({\Upsilon}{\to}B_{c}^{+}K^{-})
  &=&
   \sqrt{2}\, G_{F}\, V_{cb}\, V_{us}^{\ast}\, a_{1}\,
   f_{K}\, m_{\Upsilon}\, ({\epsilon}_{\Upsilon}{\cdot}p_{K})\,
   A_{0}^{{\Upsilon}{\to}B_{c}}
   \label{amp-bc-k}, \\
  {\cal A}({\Upsilon}{\to}B_{c}^{+}{\rho}^{-})
  &=&
   -i\,\frac{G_{F}}{\sqrt{2}}\, V_{cb}\, V_{ud}^{\ast}\,
   a_{1}\, f_{\rho}\, m_{\rho}\, \Big\{
   ({\epsilon}_{\Upsilon}{\cdot}{\epsilon}_{\rho}^{\ast})\,
   (m_{\Upsilon}+m_{B_{c}})\, A_{1}^{{\Upsilon}{\to}B_{c}}
   \nonumber \\ & & \!\!\!\!\!\!\!\!\!\!\!\!\!\!\!\!\!\!\!\!
   + ({\epsilon}_{\Upsilon}{\cdot}p_{\rho})\,
     ({\epsilon}_{\rho}^{\ast}{\cdot}p_{\Upsilon})\,
       \frac{ 2\, A_{2}^{{\Upsilon}{\to}B_{c}} }
            { m_{\Upsilon}+m_{B_{c}} }
  -i\,{\epsilon}_{{\mu}{\nu}{\alpha}{\beta}}\,
      {\epsilon}_{\Upsilon}^{\mu}\,
      {\epsilon}_{\rho}^{{\ast}{\nu}}\,
      p_{\Upsilon}^{\alpha}\,p_{\rho}^{\beta}\,
       \frac{2\, V^{{\Upsilon}{\to}B_{c}} }
            { m_{\Upsilon}+m_{B_{c}} } \Big\}
   \label{amp-bc-rho}, \\
  {\cal A}({\Upsilon}{\to}B_{c}^{+}K^{{\ast}-})
  &=&
   -i\,\frac{G_{F}}{\sqrt{2}}\, V_{cb}\, V_{us}^{\ast}\,
   a_{1}\, f_{K^{\ast}}\, m_{K^{\ast}}\, \Big\{
   ({\epsilon}_{\Upsilon}{\cdot}{\epsilon}_{K^{\ast}}^{\ast})\,
   (m_{\Upsilon}+m_{B_{c}})\, A_{1}^{{\Upsilon}{\to}B_{c}}
   \nonumber \\ & & \!\!\!\!\!\!\!\!\!\!\!\!\!\!\!\!\!\!\!\!
   + ({\epsilon}_{\Upsilon}{\cdot}p_{K^{\ast}})\,
     ({\epsilon}_{K^{\ast}}^{\ast}{\cdot}p_{\Upsilon})\,
       \frac{ 2\, A_{2}^{{\Upsilon}{\to}B_{c}} }
            { m_{\Upsilon}+m_{B_{c}} }
  -i\,{\epsilon}_{{\mu}{\nu}{\alpha}{\beta}}\,
      {\epsilon}_{\Upsilon}^{\mu}\,
      {\epsilon}_{K^{\ast}}^{{\ast}{\nu}}\,
      p_{\Upsilon}^{\alpha}\,p_{K^{\ast}}^{\beta}\,
       \frac{2\, V^{{\Upsilon}{\to}B_{c}} }
            { m_{\Upsilon}+m_{B_{c}} } \Big\}
   \label{amp-bc-kv}.
  \end{eqnarray}

  \section{Numerical results and discussion}
  \label{sec03}

  In the center-of-mass frame of Upsilons, branching ratio
  for nonleptonic ${\Upsilon}(nS)$ ${\to}$ $B_{c}M$ weak
  decays can be written as
   \begin{equation}
  {\cal B}r({\Upsilon}{\to}B_{c}M)\ =\ \frac{1}{12{\pi}}\,
   \frac{p_{\rm cm}}{m_{{\Upsilon}}^{2}{\Gamma}_{{\Upsilon}}}\,
  {\vert}{\cal A}({\Upsilon}{\to}B_{c}M){\vert}^{2}
   \label{br},
   \end{equation}
 where the momentum of final states is
   \begin{equation}
   p_{\rm cm}\ =\
   \frac{ \sqrt{ [m_{\Upsilon}^{2}-(m_{B_{c}}+m_{M})^{2}]
                 [m_{\Upsilon}^{2}-(m_{B_{c}}-m_{M})^{2}] }  }
       { 2\,m_{\Upsilon} }
   \label{pcm}.
   \end{equation}

 The input parameters,
 including the CKM Wolfenstein parameters,
 masses of $b$ and $c$ quarks, hadronic parameters including
 decay constant and Gegenbauer moment of distribution amplitudes
 in Eq.(\ref{twist}),
 are collected in Table \ref{tab:input}.
 If not specified explicitly, we will take their central
 values as the default inputs.
 Our numerical results on branching ratios for
 ${\Upsilon}(nS)$ ${\to}$ $B_{c}M$ decays are
 displayed in Table \ref{tab:br},
 where theoretical uncertainties come from the CKM parameters,
 the renormalization scale ${\mu}$ $=$ $(1{\pm}0.5)m_{b}$,
 hadronic parameters, respectively.
 For the sake of comparison, previous results of
 Refs. \cite{ijma14,adv2013} are re-evaluated with
 $a_{1}$ $=$ $1.057$.
 The following are some comments.

   \begin{table}[ht]
   \caption{Numberical values of input parameters.}
   \label{tab:input}
   \begin{ruledtabular}
   \begin{tabular}{ll}
   \multicolumn{2}{c}{Wolfenstein parameters} \\ \hline
    ${\lambda}$  $=$ $0.22537{\pm}0.00061$     \cite{pdg}
  & $A$          $=$ $0.814^{+0.023}_{-0.024}$ \cite{pdg} \\ \hline
    \multicolumn{2}{c}{masses of charm and bottom quarks} \\ \hline
    $m_{c}$ $=$ $1.67{\pm}0.07$ GeV  \cite{pdg}
  & $m_{b}$ $=$ $4.78{\pm}0.06$ GeV  \cite{pdg} \\ \hline
  \multicolumn{2}{c}{decay constants} \\ \hline
    $f_{\pi}$ $=$ $130.41{\pm}0.20$ MeV \cite{pdg}
  & $f_{K}  $ $=$ $156.2{\pm}0.7$ MeV \cite{pdg} \\
    $f_{\rho}$   $=$ $216{\pm}3$ MeV \cite{ball}
  & $f_{K^{\ast}}$ $=$ $220{\pm}5$ MeV \cite{ball} \\ \hline
  \multicolumn{2}{c}{Gegenbauer moments at scale ${\mu}$ $=$ 1 GeV} \\ \hline
    $a_{1}^{\rho}$ $=$ $0$ \cite{ball}
  & $a_{2}^{\rho}$ $=$ $0.15{\pm}0.07$ \cite{ball} \\
    $a_{1}^{K^{\ast}}$ $=$ $0.03{\pm}0.02$ \cite{ball}
  & $a_{2}^{K^{\ast}}$ $=$ $0.11{\pm}0.09$ \cite{ball} \\
    $a_{1}^{\pi}$ $=$ $0$ \cite{ball}
  & $a_{2}^{\pi}$ $=$ $0.25{\pm}0.15$ \cite{ball} \\
    $a_{1}^{K}$ $=$ $0.06{\pm}0.03$ \cite{ball}
  & $a_{2}^{K}$ $=$ $0.25{\pm}0.15$ \cite{ball}
  \end{tabular}
  \end{ruledtabular}
  \end{table}
   \begin{table}[h]
   \caption{Branching ratios for ${\Upsilon}(nS)$ ${\to}$
    $B_{c}M$ decays, where uncertainties of this work are
    from the CKM factors, scale ${\mu}$ $=$ $(1+0.5)m_{b}$,
    hadronic parameters, respectively; numbers of Ref.
    \cite{ijma14,adv2013} are evaluated with $a_{1}$ $=$ $1.057$.}
   \label{tab:br}
   \begin{ruledtabular}
   \begin{tabular}{l|c|c|c|c|c}
    final & \multicolumn{3}{c|}{${\Upsilon}(1S)$ decay}
 & ${\Upsilon}(2S)$ decay & ${\Upsilon}(3S)$ decay \\ \cline{2-6}
   states & \cite{ijma14} & \cite{adv2013}
 & this work  & this work & this work \\ \hline
   ${\cal B}r({\Upsilon}{\to}B_{c}{\rho}){\times}10^{10}$
 & $0.93$ & $0.58$
 & $1.04^{+0.07+0.05+0.03}_{-0.07-0.02-0.03}$
 & $2.47^{+0.18+0.11+0.07}_{-0.17-0.05-0.07}$
 & $3.71^{+0.26+0.17+0.10}_{-0.25-0.08-0.10}$ \\
   ${\cal B}r({\Upsilon}{\to}B_{c}{\pi}){\times}10^{11}$
 & $3.48$ & $1.43$
 & $3.39^{+0.24+0.15+0.01}_{-0.23-0.07-0.01}$
 & $8.27^{+0.59+0.37+0.03}_{-0.56-0.18-0.03}$
 & $12.40^{+0.88+0.56+0.04}_{-0.84-0.27-0.04}$ \\
   ${\cal B}r({\Upsilon}{\to}B_{c}K^{\ast}){\times}10^{12}$
 & $5.27$ & $3.12$
 & $5.26^{+0.40+0.23+0.24}_{-0.38-0.11-0.24}$
 & $12.28^{+0.94+0.54+0.57}_{-0.90-0.26-0.56}$
 & $19.09^{+1.47+0.84+0.89}_{-1.39-0.40-0.87}$ \\
   ${\cal B}r({\Upsilon}{\to}B_{c}K){\times}10^{12}$
 & $2.53$ & $1.16$
 & $2.51^{+0.19+0.11+0.03}_{-0.18-0.05-0.03}$
 & $6.18^{+0.48+0.27+0.06}_{-0.45-0.13-0.06}$
 & $9.30^{+0.72+0.41+0.09}_{-0.68-0.20-0.09}$

  \end{tabular}
  \end{ruledtabular}
  \end{table}

  (1)
  For the same final states, branching ratio of nonleptonic
  Upsilons weak decay increase with the radial quantum number
  of Upsilons below $B\bar{B}$ threshold, because of two facts,
  one is that mass of Upsilon increases with the radial quantum
  number, which results in the final phase space increases
  with the radial quantum number;
  the other is that decay width of Upsilons decreases
  with the increase of the radial quantum number of Upsilons.
  Hence, branching ratio for ${\Upsilon}(3S)$ decay
  is the largest one among nonleptonic ${\Upsilon}(1S,2S,3S)$
  weak decays into the same final $B_{c}M$ states.

  (2)
  There is a clear hierarchical relation for the same
  decaying Upsilon,
  ${\cal B}r({\Upsilon}{\to}B_{c}{\rho})$ $>$
  ${\cal B}r({\Upsilon}{\to}B_{c}{\pi})$ $>$
  ${\cal B}r({\Upsilon}{\to}B_{c}K^{\ast})$ $>$
  ${\cal B}r({\Upsilon}{\to}B_{c}K)$.
  These are two dynamical reasons.
  One is that the CKM factor
  ${\vert}V_{cb}V_{ud}^{\ast}{\vert}$ responsible for
  ${\Upsilon}$ ${\to}$ $B_{c}{\pi}$, $B_{c}{\rho}$
  decays is larger than the CKM factor
  ${\vert}V_{cb}V_{us}^{\ast}{\vert}$ responsible for
  ${\Upsilon}$ ${\to}$ $B_{c}K^{(\ast)}$ decays.
  The other is that Upsilons decay into two pseudoscalar
  mesons is suppressed by the orbital angular momentum
  with respect to Upsilons decay into $B_{c}V$ states
  with the same flavor structures.

  (3)
  The ${\Upsilon}(1S,2S,3S)$ ${\to}$ $B_{c}{\rho}$ decays
  have large branching ratio, ${\sim}$ $10^{-10}$,
  which should be sought for with high priority
  and firstly observed at the running LHC and
  forthcoming SuperKEKB.

  (4)
  There are many uncertainties.
  The first uncertainty from the CKM factors could
  be lessened with the improvement on the precision
  of the Wolfenstein parameter $A$ in the future.
  The second uncertainty from the renormalization scale
  should, in principle, be reduced by inclusion of higher
  order ${\alpha}_{s}$ corrections to hadronic matrix
  elements.
  The third uncertainty from hadronic parameters might
  be reduced with the relative ratio of branching ratios.
  For example, ingoring the kinematic effects, the relation
   \begin{equation}
   \frac{ {\cal B}r({\Upsilon}(mS){\to}B_{c}{\pi}) }
        { {\cal B}r({\Upsilon}(3S){\to}B_{c}{\pi}) }
   \ {\approx}\
   \frac{ {\cal B}r({\Upsilon}(mS){\to}B_{c}K) }
        { {\cal B}r({\Upsilon}(3S){\to}B_{c}K) }
   \ {\approx}\
   \Big(
   \frac{ A_{0}^{{\Upsilon}(mS){\to}B_{c}} }
        { A_{0}^{{\Upsilon}(3S){\to}B_{c}} }
   \Big)^{2}
   \label{rate-01},
   \end{equation}
  can be used to check various phenomenological models
  and improve our understanding on the interquark
  binding forces for heavy quarkonium.

  \section{Summary}
  \label{sec04}
  With anticipation of abundant Upsilons data sample
  at high-luminosity dedicated heavy-flavor factories,
  we studied the nonleptonic two-body bottom-changing
  ${\Upsilon}(nS)$ ${\to}$ $B_{c}M$ weak decays.
  Considering QCD radiative corrections to hadronic
  matrix elements with the QCDF approach, and using
  nonrelativistic wave functions to evaluate the
  ${\Upsilon}(nS)$ ${\to}$ $B_{c}$ transition form
  factors, we estimated the branching ratios for
  ${\Upsilon}(nS)$ ${\to}$ $B_{c}M$ weak decays.
  It is found that branching ratios for
  ${\Upsilon}(1S,2S,3S)$ ${\to}$ $B_{c}{\rho}$ decays
  is large, ${\sim}$ $10^{-10}$,
  which might be detectable at the running LHC and
  forthcoming SuperKEKB.

  \section*{Acknowledgments}
  We thank Prof. Dongsheng Du (IHEP@CAS) and Prof. Yadong Yang
  (CCNU) for helpful discussion.
  The work is supported by the National Natural Science Foundation
  of China (Grant Nos. 11475055, 11275057, U1232101 and U1332103).

  
  \end{document}